\documentclass[journal]{IEEEtran}

\usepackage{tikz}
\usepackage{siunitx}
\usetikzlibrary{arrows.meta,positioning,calc}
\usepackage{amsmath,amssymb,amsfonts}
\usepackage{mathtools}
\usepackage{graphicx}
\usepackage{booktabs}
\usepackage{cite}
\usepackage{microtype}
\usepackage[hidelinks]{hyperref}

\allowdisplaybreaks

\newcommand{\ii}{\mathrm{i}}

\newcommand{\E}{\mathbb{E}}
\newcommand{\Var}{\operatorname{Var}}
\newcommand{\atanTwo}{\operatorname{atan2}}

\title{Phase-Drift Limits and Adaptive Quadrature Readout
in Programmable Photonic Processors}

\author{Gökhan~Elmas,
        Igor~A.~Litvin,
        and~Janis~Nötzel%
\thanks{G. Elmas, I. A. Litvin, and J. Nötzel are with the
Emmy-Noether Group Theoretical Quantum Systems Design,
Technical University of Munich, Munich, Germany.}%
\thanks{Corresponding author: Gökhan Elmas
(e-mail: gokhan.elmas@tum.de).}%
}

\begin{document}

\maketitle
\begin{abstract}
Sequential acquisition of complementary optical quadratures combines
measurements associated with different underlying phases and therefore
introduces a reconstruction error when the phase evolves between
measurements. We derive the exact circular end-time errors for the two
temporal acquisition orders, cosine-first (\(C\rightarrow S\)) and
sine-first (\(S\rightarrow C\)), without assuming a stochastic model for
the phase evolution. For a phase increment \(\delta_\tau\) accumulated
over a quadrature separation \(\tau\), their small-increment limits are
\(e_{C\rightarrow S}=-\delta_\tau\sin^2\phi_0+
\mathcal{O}(\delta_\tau^2)\) and
\(e_{S\rightarrow C}=-\delta_\tau\cos^2\phi_0+
\mathcal{O}(\delta_\tau^2)\). The complementary phase dependence yields
a phase-predicted ordering rule. For phase-independent
\(Q_\tau=\E[\delta_\tau^2]\) and uniform phase coverage, the fixed-order
and adaptive mean-square errors approach \(3Q_\tau/8\) and
\((3/8-1/\pi)Q_\tau\), respectively, corresponding to an \(84.9\%\)
small-increment reduction.

We test the general theory using a fixed four-output programmable
photonic receiver and a continuous 30-min four-channel single-photon
record. Temporal acquisition is constructed from the common record by
introducing a controlled sample offset, thereby isolating the loss of
quadrature simultaneity from switching and run-to-run effects. At
\(\tau=\SI{5}{\milli\second}\), the measured first-order coefficients
follow the predicted phase dependence with correlations above \(0.998\),
while the exact nonlinear model reproduces the pointwise same-record
circular errors with correlations near \(0.992\). Applying the
phase-predicted ordering rule reduces the measured fixed-order
mean-square discrepancy from \(0.049182~\mathrm{rad}^2\) to
\(0.014010~\mathrm{rad}^2\), a \(71.51\%\) reduction; the exact theory
evaluated for the measured phase evolution predicts \(72.11\%\).

Finally, we distinguish the total phase-evolution statistic
\(Q_\tau\) from the residual conditional uncertainty \(V_\tau\) relevant
after predictable phase motion is propagated. A stale measurement with
local phase information \(J\) then contributes effective end-time
information \(J_{\rm eff}=J/(1+JV_\tau)\), providing a basis for
spatial--temporal receiver-resource tradeoffs.
\end{abstract}


\section{Introduction}
\label{sec:introduction}

Programmable photonic processors implement reconfigurable linear
transformations using interferometric meshes and tunable phase elements
\cite{Clements2016,Harris2018,Bogaerts2020,Perez2017,Carolan2015,
Taballione2019,Taballione2021}. Their operation is therefore sensitive
to relative phase fluctuations, which can occur on time scales relevant
to measurement, calibration, and feedback
\cite{Elmas2025Phase,Litvin2025Stable,Litvin2026Stabilization,
Litvin2026JointDetection}. Calibration and self-configuration methods
address transfer-matrix errors and hardware imperfections
\cite{Miller2013,Annoni2017,PerezLopez2020,Bandyopadhyay2021,
Xu2022,Markowitz2023,Zheng2024,Litvin2025Calibration,
Elmas2026Variance}, but they do not remove the distinct error that arises
when observables associated with different underlying phases are combined
in a single estimator.

For complementary quadratures \(C=\cos\phi\) and \(S=\sin\phi\),
simultaneous readout gives
\(\widehat{\phi}=\atanTwo(S,C)\). A resource-reduced receiver can instead
acquire the two quadratures sequentially. Sensitivity to temporal
variation is well established in phase-shifting interferometry
\cite{Kinnstaetter1988,Deck1998,deGroot1995,Deck2009,Cheng2018}, and
temporally interleaved in-phase and quadrature measurements have also
been considered for single-detector interferometry \cite{Roztocki2021}.
In such receivers, however, the two quantities entering the phase
reconstruction need not correspond to the same underlying phase.
The relevant receiver-design question is therefore quantitative: for a
quadrature separation \(\tau\), what phase-reconstruction error is created
solely by the loss of simultaneity, how does that error depend on the
starting phase and acquisition order, and when can a reduction in receiver
resources compensate for the resulting temporal penalty?

In this work, we develop a trajectory-level description of this mixed-time
error for complementary quadrature measurements. The treatment does not
require Brownian, Ornstein--Uhlenbeck, Gaussian, stationary, or zero-mean
phase evolution. For each realized phase increment, the two temporal
acquisition orders admit exact circular end-time errors. Their
small-increment limits reveal complementary phase sensitivities, which in
turn provide a direct rule for selecting the less drift-sensitive
acquisition order. We then test both the local and exact nonlinear
predictions using a continuous four-channel photon-count record from a
programmable photonic receiver. Finally, we distinguish the total
inter-measurement phase evolution relevant to mixed-time reconstruction
from the residual uncertainty that remains after predictable phase motion
is propagated, allowing the measurement problem to be connected to
spatial--temporal receiver-resource tradeoffs.

We take the estimation target to be the phase at the second measurement,
\begin{equation}
\phi_1=\phi_0+\delta_\tau ,
\label{eq:intro_increment}
\end{equation}
where \(\delta_\tau\) is the realized phase increment during the
quadrature separation. No probability distribution is assumed for
\(\delta_\tau\) in the trajectory-level theory. The two possible temporal
acquisition orders are
\begin{subequations}
\label{eq:intro_temporal_estimators}
\begin{align}
\widehat{\phi}_{C\rightarrow S}
&=\atanTwo(\sin\phi_1,\cos\phi_0),\\
\widehat{\phi}_{S\rightarrow C}
&=\atanTwo(\sin\phi_0,\cos\phi_1).
\end{align}
\end{subequations}
For arbitrary \(\delta_\tau\), these estimators define exact circular
end-time errors relative to \(\phi_1\). Expanding about
\(\delta_\tau=0\) gives
\begin{subequations}
\label{eq:intro_first_order}
\begin{align}
e_{C\rightarrow S}
&=-\delta_\tau\sin^2\phi_0+\mathcal{O}(\delta_\tau^2),\\
e_{S\rightarrow C}
&=-\delta_\tau\cos^2\phi_0+\mathcal{O}(\delta_\tau^2).
\end{align}
\end{subequations}
The complementary phase dependence means that one acquisition order is
least sensitive precisely where the other is most sensitive. It therefore
provides a natural phase-predicted ordering rule, while the exact nonlinear
expressions remain available when the accumulated phase increment is not
perturbatively small.

Experimentally, we test these relations with a fixed four-output
programmable photonic receiver and a continuous 30-min four-channel photon
record. Both quadratures are recorded simultaneously, and temporal
acquisition is constructed from the common record by introducing a
controlled sample offset in post-processing. This same-record construction
deliberately isolates the mixed-time mechanism from switching transients,
configuration-dependent loss, and run-to-run phase variations. It should
therefore be interpreted as a controlled experimental test of the loss of
quadrature simultaneity rather than as an end-to-end demonstration of a
physically switched sequential receiver.

At \(\tau=\SI{5}{\milli\second}\), the experimentally fitted
phase-dependent first-order coefficients follow the predicted
\(-\sin^2\phi_0\) and \(-\cos^2\phi_0\) dependences with correlations
above \(0.998\). The exact nonlinear theory also reproduces the measured
pointwise same-record circular errors with correlations near \(0.992\).
Applying the phase-predicted ordering rule sample by sample reduces the
measured same-record mean-square discrepancy from the fixed-order mean of
\(0.049182~\mathrm{rad}^2\) to \(0.014010~\mathrm{rad}^2\), corresponding
to a \(71.51\%\) reduction. Evaluating the same ordering rule with the
exact nonlinear theory for the experimentally observed phase evolution
predicts a \(72.11\%\) reduction. The close agreement shows that the
general mixed-time model accounts not only for the local phase dependence
of the error but also for its finite-increment adaptive performance.

Finally, two phase-evolution statistics must be distinguished. The direct
mixed-time reconstruction penalty depends on the total increment second
moment
\(Q_\tau=\E[\delta_\tau^2]\), which includes both predictable and
unpredictable phase evolution. By contrast, an increment-aware estimator
can first propagate the predictable component of the phase motion and is
then limited by the residual conditional uncertainty \(V_\tau\). For a
stale measurement carrying local phase information \(J\), this residual
uncertainty reduces its effective end-time information to
\begin{equation}
J_{\rm eff}
=
\frac{J}{1+JV_\tau}.
\end{equation}
This separation leads naturally to the distinct dimensionless parameters
\(\chi_Q=I_{\rm sp}Q_\tau\) and
\(\chi_V=I_{\rm sp}V_\tau\), which describe different spatial--temporal
receiver tradeoffs and should not be identified without an explicit
dynamical relation between \(Q_\tau\) and \(V_\tau\).
\section{Receiver Model and Mixed-Time Error}
\label{sec:receiver_problem}

For two coherent equal-power inputs with relative phase \(\phi\), a
linear receiver can produce four logical outputs
\begin{subequations}
\label{eq:four_output_powers}
\begin{align}
P_1&=\frac{P_0}{2}(1+\cos\phi),&
P_2&=\frac{P_0}{2}(1-\cos\phi),\\
P_3&=\frac{P_0}{2}(1+\sin\phi),&
P_4&=\frac{P_0}{2}(1-\sin\phi).
\end{align}
\end{subequations}
Balanced differences give
\begin{equation}
C=\frac{P_1-P_2}{P_1+P_2}=\cos\phi,
\qquad
S=\frac{P_3-P_4}{P_3+P_4}=\sin\phi .
\label{eq:balanced_quadratures}
\end{equation}
Unequal input powers introduce a common visibility
\(V=2\sqrt{P_aP_b}/(P_a+P_b)\), so that
\(C=V\cos\phi\) and \(S=V\sin\phi\); the phase remains recovered by
\(\atanTwo(S,C)\) provided both quadratures share the same visibility.
\begin{figure*}[t]
\centering
\resizebox{0.94\textwidth}{!}{%
\begin{tikzpicture}[
    font=\small,
    block/.style={
        draw,
        rounded corners,
        align=center,
        minimum height=0.85cm,
        minimum width=2.3cm
    },
    smallblock/.style={
        draw,
        rounded corners,
        align=center,
        minimum height=0.68cm,
        minimum width=1.65cm
    },
    output/.style={
        draw,
        rounded corners,
        align=center,
        minimum height=0.62cm,
        minimum width=2.05cm
    },
    arrow/.style={
        -{Latex[length=2.2mm]},
        thick
    },
    timeline/.style={
        thick
    }
]

% ------------------------------------------------------------
% Panel (a): logical receiver
% ------------------------------------------------------------

\node[font=\bfseries] at (-5.8,3.1)
{(a) Four-output quadrature receiver};

\node[smallblock] (ea) at (-7.0,1.8)
{$E_a$};

\node[smallblock] (eb) at (-7.0,0.4)
{$E_b$};

\node[block] (mesh) at (-4.4,1.1)
{Fixed linear\\photonic receiver};

\draw[arrow] (ea.east) -- (mesh.west |- ea.east);
\draw[arrow] (eb.east) -- (mesh.west |- eb.east);

\node[output] (p1) at (-1.5,2.5)
{$P_1\propto 1+\cos\phi$};

\node[output] (p2) at (-1.5,1.55)
{$P_2\propto 1-\cos\phi$};

\node[output] (p3) at (-1.5,0.60)
{$P_3\propto 1+\sin\phi$};

\node[output] (p4) at (-1.5,-0.35)
{$P_4\propto 1-\sin\phi$};

\draw[arrow] (mesh.east) -- (p1.west);
\draw[arrow] (mesh.east) -- (p2.west);
\draw[arrow] (mesh.east) -- (p3.west);
\draw[arrow] (mesh.east) -- (p4.west);

\node[block] (c) at (1.35,2.0)
{$C=\dfrac{P_1-P_2}{P_1+P_2}$};

\node[block] (s) at (1.35,0.1)
{$S=\dfrac{P_3-P_4}{P_3+P_4}$};

\draw[arrow] (p1.east) -- (c.west);
\draw[arrow] (p2.east) -- (c.west);
\draw[arrow] (p3.east) -- (s.west);
\draw[arrow] (p4.east) -- (s.west);

\node[block] (atan) at (4.25,1.05)
{$\widehat{\phi}=\operatorname{atan2}(S,C)$};

\draw[arrow] (c.east) -- (atan.west);
\draw[arrow] (s.east) -- (atan.west);

% ------------------------------------------------------------
% Divider
% ------------------------------------------------------------

\draw[densely dashed] (5.8,-1.0) -- (5.8,3.35);

% ------------------------------------------------------------
% Panel (b): sampling configurations
% ------------------------------------------------------------

\node[font=\bfseries] at (10.0,3.1)
{(b) Spatial and temporal sampling};

% Timeline
\draw[timeline,-{Latex[length=2mm]}]
(7.25,2.15) -- (12.65,2.15);

% Time labels placed ABOVE timeline
\node[anchor=south] at (7.25,2.22)
{$t_0$};

\node[anchor=south] at (12.65,2.22)
{$t_1=t_0+\tau$};

% Phase relation above the timeline
\node at (9.95,2.67)
{$\phi_1=\phi_0+\delta_\tau$};

% ------------------------------------------------------------
% Spatial row
% ------------------------------------------------------------

\node[anchor=east] at (7.0,1.35)
{\textbf{Spatial}};

\node[smallblock] (sp1) at (12.65,1.35)
{$C_1,\;S_1$};

\node[anchor=west,align=left] at (14.05,1.35)
{$\widehat{\phi}_{\rm sp}
=\operatorname{atan2}(S_1,C_1)$};

% ------------------------------------------------------------
% C -> S row
% ------------------------------------------------------------

\node[anchor=east] at (7.0,0.15)
{\textbf{$C\rightarrow S$}};

\node[smallblock] (cs0) at (8.0,0.15)
{$C_0$};

\node[smallblock] (cs1) at (12.65,0.15)
{$S_1$};

\draw[arrow]
(cs0.east) --
node[above] {$\tau$}
(cs1.west);

\node[anchor=west,align=left] at (14.05,0.15)
{$\widehat{\phi}_{C\rightarrow S}
=\operatorname{atan2}(S_1,C_0)$};

% ------------------------------------------------------------
% S -> C row
% ------------------------------------------------------------

\node[anchor=east] at (7.0,-1.05)
{\textbf{$S\rightarrow C$}};

\node[smallblock] (sc0) at (8.0,-1.05)
{$S_0$};

\node[smallblock] (sc1) at (12.65,-1.05)
{$C_1$};

\draw[arrow]
(sc0.east) --
node[above] {$\tau$}
(sc1.west);

\node[anchor=west,align=left] at (14.05,-1.05)
{$\widehat{\phi}_{S\rightarrow C}
=\operatorname{atan2}(S_0,C_1)$};

\end{tikzpicture}%
}
\caption{
Receiver abstraction and sampling problem.
(a) A fixed linear photonic transformation produces two complementary
cosine outputs and two complementary sine outputs. Balanced differences
remove the common optical scale and provide the quadratures \(C\) and \(S\).
(b) Spatial readout evaluates both quadratures at the same time \(t_1\),
whereas temporal readout combines quadratures acquired at times separated
by \(\tau\). The two possible temporal orders are \(C\rightarrow S\) and
\(S\rightarrow C\).
}
\label{fig:receiver_concept}
\end{figure*}

Let \(t_1=t_0+\tau\), with
\(\phi_0=\phi(t_0)\), \(\phi_1=\phi(t_1)\), and
\(\delta_\tau=\phi_1-\phi_0\).
The simultaneous estimator is
\(\widehat{\phi}_{\rm sp}=\atanTwo(\sin\phi_1,\cos\phi_1)\), whereas
the temporal estimators are those in
Eq.~\eqref{eq:intro_temporal_estimators}. Circular error is defined
relative to \(\phi_1\),
\begin{equation}
e_r=\operatorname{Arg}
\!\left[e^{i(\widehat{\phi}_r-\phi_1)}\right].
\label{eq:circular_error}
\end{equation}

\subsection{Trajectory-level error}
\label{sec:fixed_order}

Expanding the two temporal estimators about \(\delta_\tau=0\) gives
\begin{subequations}
\label{eq:fixed_order_pair}
\begin{align}
e_{C\rightarrow S}
&=-\delta_\tau\sin^2\phi_0+\mathcal{O}(\delta_\tau^2),\\
e_{S\rightarrow C}
&=-\delta_\tau\cos^2\phi_0+\mathcal{O}(\delta_\tau^2).
\end{align}
\end{subequations}
Thus, one order is least sensitive exactly where the other is most
sensitive. For finite increments, Eqs.~\eqref{eq:intro_temporal_estimators}
and \eqref{eq:circular_error} define the exact nonlinear errors
\(e_{C\rightarrow S}(\phi_0,\delta_\tau)\) and
\(e_{S\rightarrow C}(\phi_0,\delta_\tau)\) without approximation. The
exact circular-error expressions and the first- and second-order
expansions are derived in Appendix~\ref{app:mixed_time_derivation}.

\subsection{Mean-square penalty}

For arbitrary phase dynamics, the exact fixed-order MSEs are
\begin{subequations}
\label{eq:general_exact_fixed_mse}
\begin{align}
D_{C\rightarrow S}^{\rm exact}
&=
\E\!\left[e_{C\rightarrow S}(\phi_0,\delta_\tau)^2\right],\\
D_{S\rightarrow C}^{\rm exact}
&=
\E\!\left[e_{S\rightarrow C}(\phi_0,\delta_\tau)^2\right].
\end{align}
\end{subequations}
These expressions are general with respect to the joint statistics of
\(\phi_0\) and \(\delta_\tau\). In the small-increment regime, they
reduce to
\begin{subequations}
\label{eq:general_fixed_mse}
\begin{align}
D_{C\rightarrow S}
&\simeq \E[\delta_\tau^2\sin^4\phi_0],\\
D_{S\rightarrow C}
&\simeq \E[\delta_\tau^2\cos^4\phi_0].
\end{align}
\end{subequations}
The leading-order expressions still allow the increment strength to
depend on the starting phase. Defining
\begin{equation}
Q_\tau=\E[\delta_\tau^2],
\label{eq:unconditional_increment_second_moment}
\end{equation}
does not impose \(\E[\delta_\tau]=0\): a systematic or predictable
increment contributes through its squared mean. If the increment second
moment is independent of \(\phi_0\) and the starting phase is uniform,
\(\E[\sin^4\phi]=\E[\cos^4\phi]=3/8\), and
\begin{equation}
D_{\rm fixed}\simeq\frac{3}{8}Q_\tau .
\label{eq:fixed_mse_uniform_limit}
\end{equation}
The general joint-statistics form and the uniform-phase averages are
derived in Appendix~\ref{app:mse_adaptive_derivation}.

\section{Phase-Predicted Adaptive Ordering}
\label{sec:adaptive_ordering}

Adaptive phase measurements use prior information to modify the
measurement configuration
\cite{Wiseman1995,BerryWiseman2000,Armen2002,Yonezawa2012,Berni2015}.
Here the observables are unchanged; only their temporal order is
selected. From Eq.~\eqref{eq:fixed_order_pair}, the leading squared
errors are proportional to
\(\delta_\tau^2\sin^4\phi_0\) and
\(\delta_\tau^2\cos^4\phi_0\). The ideal rule is therefore
\begin{equation}
r^\star(\phi_0)=
\begin{cases}
C\rightarrow S,& |\sin\phi_0|\le |\cos\phi_0|,\\
S\rightarrow C,& |\sin\phi_0|>|\cos\phi_0|.
\end{cases}
\label{eq:adaptive_rule}
\end{equation}
For arbitrary increments, the performance of this same
phase-predicted rule is evaluated with the exact nonlinear errors:
\begin{equation}
D_{\rm ad}^{\rm exact}
=
\E\!\left[
e_{r^\star(\phi_0)}(\phi_0,\delta_\tau)^2
\right].
\label{eq:adaptive_exact_mse_general}
\end{equation}
Using
\begin{equation}
\overline D_{\rm fixed}^{\rm exact}
=
\frac{
D_{C\rightarrow S}^{\rm exact}
+
D_{S\rightarrow C}^{\rm exact}
}{2},
\label{eq:exact_fixed_mean_general}
\end{equation}
the exact nonlinear reduction achieved by the rule is
\begin{equation}
R_{\rm ad}^{\rm exact}
=
1-
\frac{D_{\rm ad}^{\rm exact}}
{\overline D_{\rm fixed}^{\rm exact}}.
\label{eq:exact_adaptive_reduction_general}
\end{equation}
These expressions apply to any joint distribution of
\((\phi_0,\delta_\tau)\); the resulting numerical reduction therefore
depends on the phase dynamics being considered.

In the small-increment regime,
\begin{equation}
D_{\rm ad}
\simeq
\E\!\left[
\delta_\tau^2
\min(\sin^4\phi_0,\cos^4\phi_0)
\right].
\label{eq:adaptive_mse_general}
\end{equation}
For phase-independent \(Q_\tau\) and uniform phase coverage,
\begin{equation}
D_{\rm ad}
\simeq
\left(\frac{3}{8}-\frac{1}{\pi}\right)Q_\tau ,
\label{eq:adaptive_uniform_mse}
\end{equation}
which is \(15.1\%\) of the fixed-order MSE, i.e. an \(84.9\%\)
asymptotic reduction. This closed-form value is the
uniform-phase, small-increment limit of the more general exact
performance measure in Eq.~\eqref{eq:exact_adaptive_reduction_general};
it is not a universal finite-increment percentage. The phase-average
leading to \(3/8-1/\pi\) is evaluated explicitly in
Appendix~\ref{app:mse_adaptive_derivation}.

With a predictor \(\widetilde{\phi}_0\), Eq.~\eqref{eq:adaptive_rule}
is evaluated at \(\widetilde{\phi}_0\). Errors are most consequential
near the switching boundaries
\(\phi_0=\pi/4+m\pi/2\); away from them, moderate prediction error
need not change the selected order. We use both ideal and one-cycle
stale predictors experimentally below.

\section{Experimental Implementation and Analysis}
\label{sec:experiment}

The experiment uses an eight-mode Si$_3$N$_4$ programmable processor
(QuiX Quantum), related to the universal processors reported in
Refs.~\cite{Taballione2019,Taballione2021}. A Thorlabs MX40C source
near \(1550~\mathrm{nm}\) is divided into two coherent branches and
injected into inputs I3 and I8. The processor is programmed once and
held fixed during acquisition. Four logical outputs are recorded with
AUREA SPD\_OEM\_NIR detectors and a ChronoXea four-channel TDC; the
logical-to-physical mapping is listed in
Table~\ref{tab:logical_physical_mapping}.

\begin{table}[t]
\centering
\caption{Logical and physical receiver outputs.}
\label{tab:logical_physical_mapping}
\begin{tabular}{ccc}
\toprule
Logical output & Processor output & Ideal dependence \\
\midrule
\(P_1\) & O3 & \(1+\cos\phi\) \\
\(P_2\) & O5 & \(1-\cos\phi\) \\
\(P_3\) & O8 & \(1+\sin\phi\) \\
\(P_4\) & O7 & \(1-\sin\phi\) \\
\bottomrule
\end{tabular}
\end{table}

\begin{figure*}[t]
\centering
\resizebox{0.96\textwidth}{!}{%
\begin{tikzpicture}[
    font=\small,
    block/.style={
        draw,
        rounded corners,
        align=center,
        minimum height=0.95cm,
        minimum width=2.45cm
    },
    smallblock/.style={
        draw,
        rounded corners,
        align=center,
        minimum height=0.72cm,
        minimum width=1.90cm
    },
    detector/.style={
        draw,
        rounded corners,
        align=center,
        minimum height=0.65cm,
        minimum width=1.35cm
    },
    arrow/.style={
        -{Latex[length=2.2mm]},
        thick
    },
    bus/.style={
        thick
    }
]

% ------------------------------------------------------------
% Optical input
% ------------------------------------------------------------

\node[block] (source) at (-7.3,0)
{Coherent\\optical source};

\node[block] (split) at (-4.5,0)
{Two coherent\\input branches};

\node[block] (chip) at (-1.4,0)
{Fixed programmable\\photonic receiver};

\draw[arrow] (source.east) -- (split.west);
\draw[arrow] (split.east) -- (chip.west);

% ------------------------------------------------------------
% Four logical outputs
% ------------------------------------------------------------

\node[smallblock] (p1) at (1.8,2.25)
{\(P_1\)\\\(+\cos\)};

\node[smallblock] (p2) at (1.8,0.75)
{\(P_2\)\\\(-\cos\)};

\node[smallblock] (p3) at (1.8,-0.75)
{\(P_3\)\\\(+\sin\)};

\node[smallblock] (p4) at (1.8,-2.25)
{\(P_4\)\\\(-\sin\)};

\draw[arrow] (chip.east) -- (p1.west);
\draw[arrow] (chip.east) -- (p2.west);
\draw[arrow] (chip.east) -- (p3.west);
\draw[arrow] (chip.east) -- (p4.west);

% ------------------------------------------------------------
% Detectors
% ------------------------------------------------------------

\node[detector] (d1) at (4.2,2.25) {SPD 1};
\node[detector] (d2) at (4.2,0.75) {SPD 2};
\node[detector] (d3) at (4.2,-0.75) {SPD 3};
\node[detector] (d4) at (4.2,-2.25) {SPD 4};

\draw[arrow] (p1.east) -- (d1.west);
\draw[arrow] (p2.east) -- (d2.west);
\draw[arrow] (p3.east) -- (d3.west);
\draw[arrow] (p4.east) -- (d4.west);

% ------------------------------------------------------------
% SPD collection bus
% ------------------------------------------------------------

\coordinate (bus1) at (5.35,2.25);
\coordinate (bus2) at (5.35,0.75);
\coordinate (bus3) at (5.35,-0.75);
\coordinate (bus4) at (5.35,-2.25);
\coordinate (busmid) at (5.35,0);

\draw[bus] (bus1) -- (bus4);

\draw[arrow] (d1.east) -- (bus1);
\draw[arrow] (d2.east) -- (bus2);
\draw[arrow] (d3.east) -- (bus3);
\draw[arrow] (d4.east) -- (bus4);

% ------------------------------------------------------------
% Four-channel TDC
% ------------------------------------------------------------

\node[
    block,
    minimum width=2.85cm
] (tdc) at (7.7,0)
{Four-channel\\time-to-digital converter};

\draw[arrow] (busmid) -- (tdc.west);

% ------------------------------------------------------------
% Balanced quadrature reconstruction
% ------------------------------------------------------------

\node[
    block,
    minimum width=2.85cm
] (quad) at (11.5,0)
{Balanced quadratures\\
\(C(t),\,S(t)\)};

\draw[arrow] (tdc.east) -- (quad.west);

% ------------------------------------------------------------
% Spatial and temporal analysis blocks
% ------------------------------------------------------------

\node[
    block,
    minimum width=2.55cm
] (spatial) at (15.1,1.35)
{Spatial\\
\(\widehat{\phi}_{\rm sp}(t)\)};

\node[
    block,
    minimum width=2.55cm
] (temporal) at (15.1,-1.35)
{Temporal\\
\(\widehat{\phi}_{C\rightarrow S}\),\\
\(\widehat{\phi}_{S\rightarrow C}\)};

% ------------------------------------------------------------
% Clean split from quadratures
% ------------------------------------------------------------

\coordinate (splitmid) at (13.4,0);
\coordinate (splittop) at (13.4,1.35);
\coordinate (splitbottom) at (13.4,-1.35);

% line from quadrature block to split bus
\draw[bus] (quad.east) -- (splitmid);

% vertical split bus
\draw[bus] (splitbottom) -- (splittop);

% arrows into analysis blocks
\draw[arrow] (splittop) -- (spatial.west);
\draw[arrow] (splitbottom) -- (temporal.west);

% ------------------------------------------------------------
% Same-record comparison block
% ------------------------------------------------------------

\node[
    block,
    minimum width=2.65cm
] (compare) at (18.85,0)
{Same-record\\error comparison};

% ------------------------------------------------------------
% Clean merge from analysis branches
% ------------------------------------------------------------

\coordinate (mergetop) at (17.0,1.35);
\coordinate (mergebottom) at (17.0,-1.35);
\coordinate (mergemid) at (17.0,0);

% lines from analysis blocks to merge bus
\draw[bus] (spatial.east) -- (mergetop);
\draw[bus] (temporal.east) -- (mergebottom);

% vertical merge bus
\draw[bus] (mergebottom) -- (mergetop);

% single arrow to comparison block
\draw[arrow] (mergemid) -- (compare.west);

% ------------------------------------------------------------
% Post-processing annotation
% ------------------------------------------------------------

\node[
    align=center
] at (13.5,-2.75)
{temporal delay \(\tau\) introduced only in post-processing};

\end{tikzpicture}%
}
\caption{
Experimental architecture for same-record spatial--temporal comparison.
A fixed programmable photonic transformation produces complementary
cosine and sine output pairs, which are detected simultaneously by four
single-photon detectors. The complete four-channel photon record is
acquired without reconfiguring the photonic processor. Spatial readout
uses simultaneous quadratures, whereas temporal readout is emulated by
combining measured quadratures separated by a controlled delay
\(\tau\). Consequently, both estimates are derived from the same
experimental phase trajectory.
}
\label{fig:experimental_architecture}
\end{figure*}

The final record has duration \(T_{\rm rec}=\SI{1800}{\second}\).
Detectors were operated at \(10\%\) efficiency and
\(\tau_{\rm d}=\SI{30}{\micro\second}\) deadtime. Counts were corrected
with the nonparalyzable deadtime model and independently measured dark
rates were subtracted. Native \(\SI{0.5}{\milli\second}\) bins were
aggregated to \(\Delta t=\SI{5}{\milli\second}\) for the principal
analysis, yielding \(360\,000\) analysis bins. The same processing is
applied identically to all four channels.

The measured quadratures and simultaneous phase are
\begin{subequations}
\label{eq:experimental_quadratures}
\begin{align}
C_k&=\frac{R_1[k]-R_2[k]}{R_1[k]+R_2[k]},\\
S_k&=\frac{R_3[k]-R_4[k]}{R_3[k]+R_4[k]},
\end{align}
\end{subequations}
\begin{equation}
\widehat{\phi}_{\rm sp}[k]=\atanTwo(S_k,C_k).
\label{eq:experimental_spatial_phase}
\end{equation}
For \(\tau=m\Delta t\), same-record temporal readout is
\begin{subequations}
\label{eq:same_record_temporal}
\begin{align}
\widehat{\phi}_{C\rightarrow S}[k;m]
&=\atanTwo(S_{k+m},C_k),\\
\widehat{\phi}_{S\rightarrow C}[k;m]
&=\atanTwo(S_k,C_{k+m}),
\end{align}
\end{subequations}
and both are compared with the simultaneous end-time reference
\(\widehat{\phi}_{\rm sp}[k+m]\) using circular errors. The empirical
increment second moment is obtained directly from the unwrapped
simultaneous phase,
\begin{equation}
\widehat{Q}_\tau
=
\frac{1}{N_\tau}
\sum_k
[\widetilde{\phi}_{\rm sp}(k+m)-\widetilde{\phi}_{\rm sp}(k)]^2 .
\label{eq:empirical_Q}
\end{equation}
No increment distribution or scaling law in \(\tau\) is assumed.
Because samples belong to one continuous trajectory, reported
uncertainty intervals use a block bootstrap with \(\SI{5}{\second}\)
blocks; block lengths from 1--20 s give consistent conclusions.

The same-record construction deliberately does not emulate switching
transients or configuration-dependent loss. It tests the mixed-time
mechanism under a common experimental trajectory; a physically
sequential implementation may add further hardware-specific errors.

\section{Experimental Results}
\label{sec:experimental_validation}

The quadrature locus is close to the ideal unit circle. The median
radius is \(0.98539\), and the RMS deviation from unity is \(0.06978\).
This supports the balanced-quadrature model without projecting the
measured data onto the unit circle.

\begin{figure}[t]
\centering
\includegraphics[
    width=0.48\textwidth
]{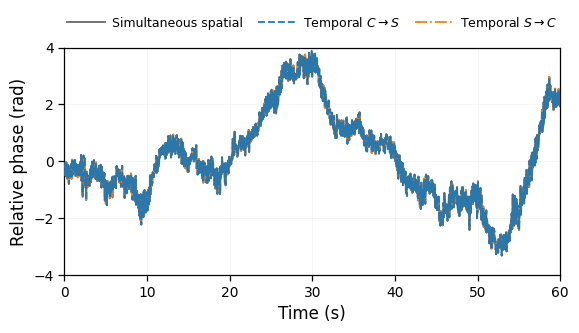}
\caption{
Representative 60-s segment of the reconstructed phase trajectory
from the 30-min experimental record. The simultaneous spatial
estimate and the two temporally separated reconstructions,
\(C\rightarrow S\) and \(S\rightarrow C\), are shown for a
quadrature separation of \(\tau=\SI{5}{\milli\second}\).
The temporal estimates closely follow the simultaneous phase
trajectory.
}
\label{fig:phase_trace_60s}
\end{figure}

Figure~\ref{fig:phase_trace_60s} shows a representative 60-s interval
from the 30-min experimental record. The simultaneous spatial
reconstruction and both temporal reconstruction orders follow the same
underlying phase evolution over the displayed interval. At this scale,
the temporal estimates largely overlap the simultaneous reference,
showing that all three readout strategies recover the same overall
phase trajectory.

The close overlap in Fig.~\ref{fig:phase_trace_60s} does not imply that
temporal separation is error free. The relevant differences occur
locally on the scale of the phase increment accumulated during the
quadrature separation. We therefore quantify the temporal-readout
penalty through the circular discrepancy relative to the simultaneous
end-time phase.

The principal analysis uses
\(\tau=\Delta t=\SI{5}{\milli\second}\), corresponding to \(m=1\).
The reference-relative same-record discrepancies are
\begin{align}
D_{C\rightarrow S}^{\rm(exp)}
&=
0.049399~\mathrm{rad}^2,
\nonumber\\
D_{S\rightarrow C}^{\rm(exp)}
&=
0.048965~\mathrm{rad}^2.
\label{eq:experimental_fixed_mse}
\end{align}
The corresponding 95\% block-bootstrap intervals are
\([0.04710,0.05180]~\mathrm{rad}^2\) and
\([0.04616,0.05178]~\mathrm{rad}^2\), respectively.
These quantities are mean-square discrepancies relative to the
simultaneous end-time estimate
\(\widehat{\phi}_{\rm sp}[k+1]\), rather than absolute errors relative
to an independently known physical phase.

\subsection{Phase-dependent and nonlinear validation}

We first test the local first-order error laws. Samples satisfying
\[
0.03<|\delta_\tau|<0.2~\mathrm{rad}
\]
are grouped according to their starting phase \(\phi_0\). Within each
phase bin, the measured mixed-time error is fitted to the
zero-intercept relation
\begin{equation}
e\simeq a(\phi_0)\delta_\tau.
\end{equation}
Thus, each phase bin provides one experimentally fitted coefficient
\(a(\phi_0)\). These fitted coefficients are compared with the
theoretical predictions
\(-\sin^2\phi_0\) for \(C\rightarrow S\) and
\(-\cos^2\phi_0\) for \(S\rightarrow C\).

The resulting correlations are
\begin{align}
r_{C\rightarrow S}
&=
0.99819,
\nonumber\\
r_{S\rightarrow C}
&=
0.99808,
\label{eq:first_order_correlations}
\end{align}
with coefficient RMSEs \(0.02182\) and \(0.02250\), respectively.

\begin{figure}[t]
\centering
\includegraphics[
    width=0.48\textwidth
]{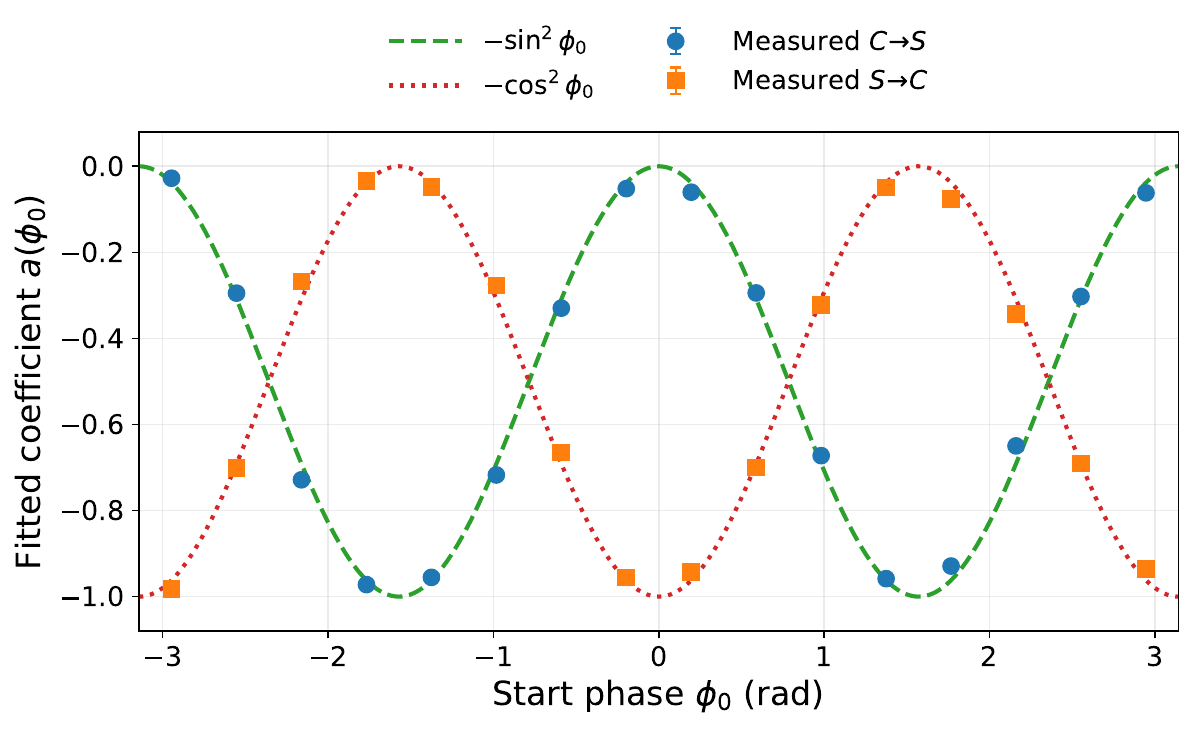}
\caption{
Direct experimental test of the phase-dependent first-order temporal
error laws at \(\tau=\SI{5}{\milli\second}\).
Samples in the local regime
\(0.03<|\delta_\tau|<0.2~\mathrm{rad}\)
are grouped by starting phase.
Within each phase bin, the relation
\(e\simeq a(\phi_0)\delta_\tau\)
is fitted with zero intercept.
The resulting fitted coefficients closely follow the analytical
predictions \(-\sin^2\phi_0\) for \(C\rightarrow S\) and
\(-\cos^2\phi_0\) for \(S\rightarrow C\).
Error bars denote the standard errors of the fitted slopes.
}
\label{fig:exp_first_order_coefficients}
\end{figure}

We next test the exact nonlinear theory without restricting the phase
increment to the local small-increment regime. For each measured phase
pair
\[
\left(
\widehat{\phi}_{\rm sp}[k],
\widehat{\phi}_{\rm sp}[k+1]
\right),
\]
the exact \(C\rightarrow S\) and \(S\rightarrow C\) phase estimates
are calculated from Eq.~\eqref{eq:intro_temporal_estimators}, and their
circular end-time errors are obtained using
Eq.~\eqref{eq:circular_error}.

Averaging the squared exact theoretical errors gives
\begin{align}
D_{C\rightarrow S}^{\rm(th)}
&=
0.047973~\mathrm{rad}^2,
\nonumber\\
D_{S\rightarrow C}^{\rm(th)}
&=
0.047879~\mathrm{rad}^2.
\label{eq:exact_fixed_validation}
\end{align}
These values are close to the corresponding experimental discrepancies
in Eq.~\eqref{eq:experimental_fixed_mse}.

The comparison is also performed sample by sample. For each
acquisition order, the exact theoretical circular-error sequence is
compared directly with the experimentally reconstructed same-record
circular-error sequence. The Pearson correlations between measured and
theoretical errors are approximately \(0.992\) for both acquisition
orders. Thus, the agreement is not limited to the average MSE: the
exact nonlinear theory also reproduces the temporal variation of the
individual mixed-time errors.

\begin{figure}[t]
\centering
\includegraphics[
    width=0.48\textwidth
]{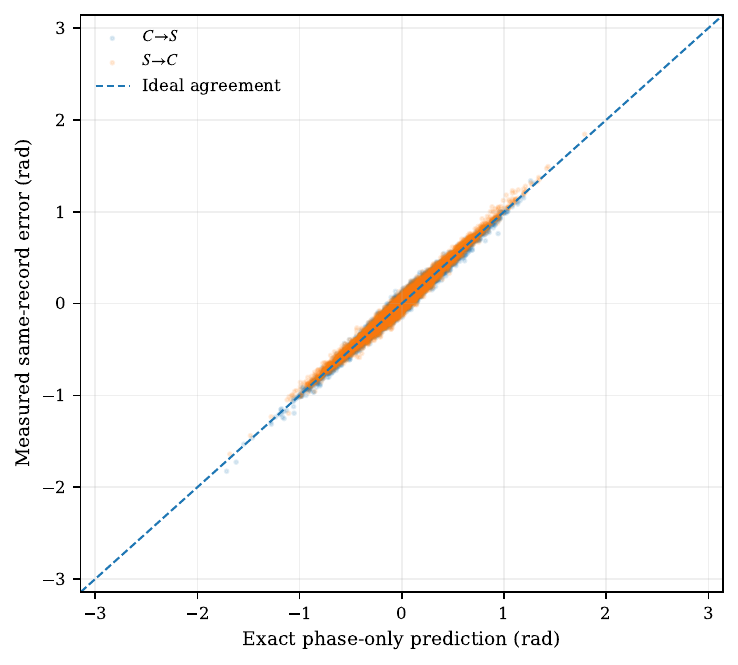}
\caption{
Pointwise comparison between the experimentally reconstructed
same-record temporal error and the exact phase-only mixed-time
prediction at \(\tau=\SI{5}{\milli\second}\) for the 30-min
acquisition.
Each point compares the measured and theoretically predicted circular
error for the same temporal sample pair.
The dashed diagonal denotes equality.
The strong correlations for both acquisition orders show that the
exact mixed-time theory reproduces both the measured error trajectory
and its mean-square magnitude.
}
\label{fig:exp_exact_nonlinear_validation}
\end{figure}

\begin{figure}[t]
\centering
\includegraphics[
    width=0.48\textwidth
]{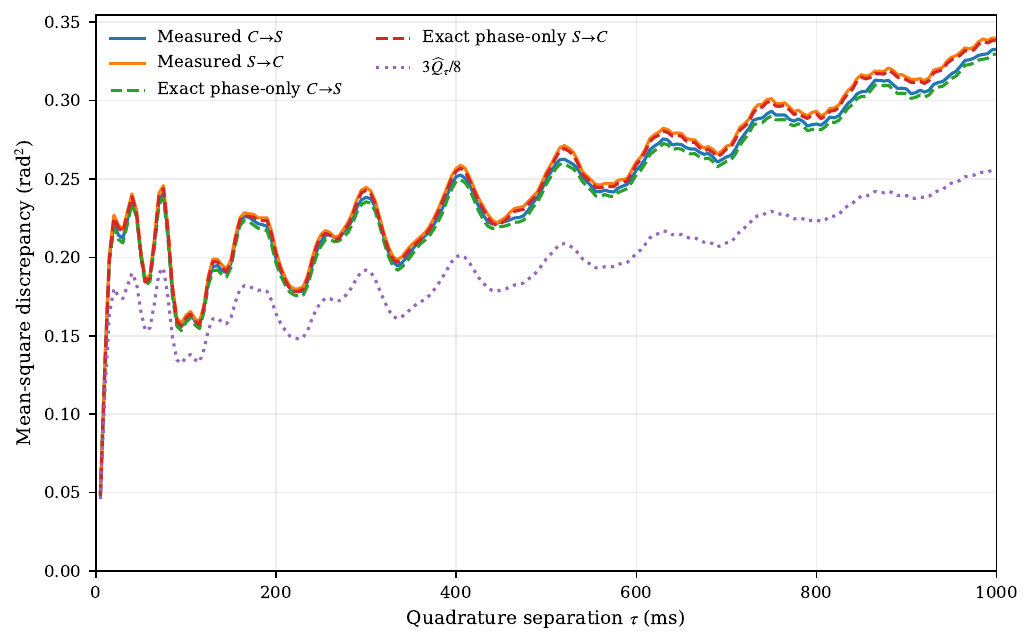}
\caption{
Same-record end-time temporal mean-square discrepancy versus
quadrature separation \(\tau\).
The two fixed acquisition orders are compared with the exact nonlinear
phase-only mixed-time predictions and the leading uniform-phase result
\(3\widehat{Q}_\tau/8\).
The exact nonlinear theory closely follows the experimental temporal
penalty over the displayed delay range, whereas the first-order result
is most accurate when the accumulated phase increment is small.
}
\label{fig:exp_mse_delay}
\end{figure}

The temporal penalty increases with quadrature separation and vanishes
at zero delay, as required by the theory.
Figure~\ref{fig:exp_mse_delay} compares the measured delay dependence
with both the exact nonlinear prediction and the leading
small-increment expression.
No Brownian or other prescribed scaling law in \(\tau\) is imposed.

\subsection{Adaptive ordering}

At \(\tau=\SI{5}{\milli\second}\), the measured fourth-order phase
moments are
\begin{align*}
\E[\sin^4\phi]
&=
0.38218,\\
\E[\cos^4\phi]
&=
0.36661.
\end{align*}
The measured adaptive coefficient is
\begin{equation}
\E\!\left[
\min(\sin^4\phi,\cos^4\phi)
\right]
=
0.05718.
\end{equation}

For uniform phase coverage, the corresponding analytical moments are
\begin{align*}
\E[\sin^4\phi]
&=
\frac{3}{8},\\
\E[\cos^4\phi]
&=
\frac{3}{8},
\end{align*}
while the adaptive coefficient is
\begin{equation}
\E\!\left[
\min(\sin^4\phi,\cos^4\phi)
\right]
=
\frac{3}{8}
-
\frac{1}{\pi}
=
0.05669.
\end{equation}

Thus, the measured phase coverage is close to the uniform-phase
condition used in deriving the closed-form small-increment result.
Substituting the measured phase moments into the leading-order
expression while retaining phase-independent \(Q_\tau\) gives an
\(84.73\%\) reduction, close to the asymptotic \(84.9\%\).
Therefore, nonuniform phase coverage alone does not explain the lower
finite-increment reduction observed experimentally.

The mean fixed-order experimental discrepancy is
\begin{align}
\overline{D}_{\rm fixed}^{\rm(exp)}
&=
\frac{
D_{C\rightarrow S}^{\rm(exp)}
+
D_{S\rightarrow C}^{\rm(exp)}
}{2}
\nonumber\\
&=
0.049182~\mathrm{rad}^2.
\label{eq:experimental_fixed_mean}
\end{align}

Applying the phase-predicted ordering rule of
Eq.~\eqref{eq:adaptive_rule} sample by sample gives
\begin{equation}
D_{\rm ad}^{\rm(exp)}
=
0.014010~\mathrm{rad}^2.
\label{eq:experimental_adaptive_mse}
\end{equation}
The measured adaptive reduction is therefore
\begin{align}
R_{\rm ad}^{\rm(exp)}
&=
1-
\frac{
D_{\rm ad}^{\rm(exp)}
}{
\overline{D}_{\rm fixed}^{\rm(exp)}
}
\nonumber\\
&=
71.51\%.
\label{eq:experimental_adaptive_reduction}
\end{align}

We then evaluate the general exact theory for the same measured phase
evolution. For \(\tau=m\Delta t\), define
\begin{align}
\phi_{0,k}
&=
\widehat{\phi}_{\rm sp}[k],
\nonumber\\
\phi_{1,k}
&=
\widehat{\phi}_{\rm sp}[k+m].
\label{eq:sample_phase_pair}
\end{align}

For each sample pair, the exact theoretical circular errors are
obtained from Eqs.~\eqref{eq:intro_temporal_estimators} and
\eqref{eq:circular_error}. Defining
\[
r_k^\star=r^\star(\phi_{0,k}),
\]
the exact adaptive theoretical MSE is evaluated as
\begin{equation}
D_{\rm ad}^{\rm(th)}
=
\frac{1}{N_\tau}
\sum_k
\left(
e_{r_k^\star,k}^{\rm(th)}
\right)^2.
\label{eq:exact_adaptive_mse_definition}
\end{equation}

At \(\tau=\SI{5}{\milli\second}\), the exact fixed-order theory gives
\begin{align}
D_{C\rightarrow S}^{\rm(th)}
&=
0.047973~\mathrm{rad}^2,
\nonumber\\
D_{S\rightarrow C}^{\rm(th)}
&=
0.047879~\mathrm{rad}^2.
\label{eq:exact_fixed_mse_numeric}
\end{align}

Their mean is
\begin{align}
\overline D_{\rm fixed}^{\rm(th)}
&=
\frac{
0.047973+0.047879
}{2}
\nonumber\\
&=
0.047926~\mathrm{rad}^2.
\label{eq:exact_fixed_mean_numeric}
\end{align}

Applying the same phase-predicted ordering rule to the exact nonlinear
errors gives
\begin{equation}
D_{\rm ad}^{\rm(th)}
=
0.013366~\mathrm{rad}^2.
\label{eq:exact_adaptive_mse}
\end{equation}

The corresponding exact-theory reduction is
\begin{align}
R_{\rm ad}^{\rm(th)}
&=
1-
\frac{
D_{\rm ad}^{\rm(th)}
}{
\overline D_{\rm fixed}^{\rm(th)}
}
\nonumber\\
&=
1-
\frac{0.013366}{0.047926}
\nonumber\\
&=
0.7211
\nonumber\\
&=
72.11\%.
\label{eq:exact_adaptive_reduction}
\end{align}

Thus, the general exact theory evaluated for the phase evolution present
in the experiment predicts a \(72.11\%\) reduction, compared with the
measured \(71.51\%\).
The difference is only \(0.60\) percentage points.
The \(84.9\%\) result is therefore interpreted as the uniform-phase,
small-increment analytical limit, whereas the exact nonlinear theory
accounts for the finite phase increments encountered experimentally.

Finally, a one-cycle-stale predictor has RMS phase error
\(0.3499~\mathrm{rad}\), agrees with the ideal ordering decision in
\(82.80\%\) of the samples, and still reduces the discrepancy by
\(47.17\%\).
This shows that a substantial adaptive benefit remains even when the
ordering decision is based on imperfect phase information.

\section{Increment-Aware End-Time Estimation}
\label{sec:increment_aware}

Having experimentally validated the mixed-time error mechanism and
phase-predicted ordering in Sec.~\ref{sec:experimental_validation}, we
now consider a theoretical extension in which predictable phase
evolution is explicitly propagated before an earlier measurement is
reused. Unlike the mixed-time analysis above, which treats the realized
increment directly, this estimator down-weights only the unresolved
part of the inter-measurement evolution. Standard local Gaussian
estimation and state-space arguments
\cite{Kay1993,Sarkka2013} provide a convenient closed-form description.

For a noisy scalar cosine or sine quadrature linearized around a phase
predictor \(\bar{\phi}\), the corresponding local phase-information
quantities are
\begin{align}
J_C
&=
\frac{\sin^2\bar{\phi}}{v_C},
\nonumber\\
J_S
&=
\frac{\cos^2\bar{\phi}}{v_S},
\label{eq:local_information}
\end{align}
where \(v_C\) and \(v_S\) are the local quadrature-noise variances.

Let \(\mathcal H_0\) denote the information available before the
end-time measurement. We decompose the realized phase increment as
\begin{align}
m_\tau
&=
\E[\delta_\tau\mid\mathcal H_0],
\nonumber\\
w_\tau
&=
\delta_\tau-m_\tau,
\nonumber\\
V_\tau
&=
\Var(w_\tau\mid\mathcal H_0).
\label{eq:local_increment_model}
\end{align}
Here \(m_\tau\) is the predictable component of the phase increment,
while \(w_\tau\) is the remaining prediction residual. By construction,
\[
\E[w_\tau\mid\mathcal H_0]=0.
\]
This condition applies only to the residual \(w_\tau\); no zero-mean
assumption is imposed on the physical increment \(\delta_\tau\).

Consider an earlier local phase observation with information \(J\).
After propagating that observation by the predicted increment
\(m_\tau\), its end-time variance is
\begin{equation}
J^{-1}+V_\tau.
\end{equation}
Its effective end-time information is therefore
\begin{equation}
\boxed{
J_{\rm eff}
=
\frac{1}{J^{-1}+V_\tau}
=
\frac{J}{1+JV_\tau}.
}
\label{eq:general_effective_information}
\end{equation}

This expression has a direct interpretation. If the inter-measurement
evolution is predicted perfectly, \(V_\tau=0\), then
\(J_{\rm eff}=J\) and the earlier measurement loses no information.
As the unresolved prediction uncertainty increases,
\(J_{\rm eff}\) decreases.

For \(C\rightarrow S\), the cosine observation is the earlier
measurement and the sine observation is fresh at the end time.
The corresponding local MSE is
\begin{equation}
D_{C\rightarrow S}^{\rm IA}
=
\left[
J_S
+
\frac{J_C}{1+J_CV_\tau}
\right]^{-1},
\label{eq:ia_cs_mse}
\end{equation}
where ``IA'' denotes increment-aware estimation. For
\(S\rightarrow C\), the roles of \(C\) and \(S\) are exchanged.

The information lost by delaying a measurement of information \(J\) is
\begin{equation}
\Delta J
=
J-J_{\rm eff}
=
\frac{J^2V_\tau}{1+JV_\tau}.
\label{eq:information_loss}
\end{equation}
Since \(\Delta J\) increases with \(J\), delaying a more informative
quadrature incurs the larger penalty. The information argument
therefore reproduces the same ordering principle as
Eq.~\eqref{eq:adaptive_rule}: the less informative quadrature should be
measured first, leaving the more informative quadrature fresh at the
end time.

The distinction between \(Q_\tau\) and \(V_\tau\) is essential.
Conditioned on \(\mathcal H_0\),
\begin{equation}
\E[\delta_\tau^2\mid\mathcal H_0]
=
m_\tau^2+V_\tau.
\label{eq:conditional_second_moment}
\end{equation}
Averaging over predictor states gives
\begin{equation}
Q_\tau
=
\E[\delta_\tau^2]
=
\E[m_\tau^2+V_\tau].
\label{eq:Q_V_relation}
\end{equation}
Thus \(Q_\tau\) quantifies the total mean-square phase evolution,
including predictable motion, whereas \(V_\tau\) quantifies only the
uncertainty that remains after prediction.

The present experiment does not independently estimate \(V_\tau\) or
implement the increment-aware estimator of
Eq.~\eqref{eq:ia_cs_mse}. The results in this section should therefore
be interpreted as a theoretical extension of the experimentally
validated mixed-time framework rather than as an additional
experimental validation. The local linearization, information update,
and inverse-variance combination are derived in
Appendix~\ref{app:increment_aware_derivation}. A fully nonlinear
implementation could instead propagate a circular state distribution
\cite{Sarkka2013}; the local model is used here to obtain the
closed-form information law.

\section{Spatial--Temporal Resource Tradeoff}
\label{sec:resource}

For an ideal complementary Poisson pair with mean signal-photon number
\(\Lambda\) and total efficiency \(\eta\), the Fisher information is
phase independent and equals the detected signal-photon number,
\(I=\eta\Lambda\). We use this as an ideal resource benchmark rather
than an exact likelihood model for the nonideal experimental receiver.

Let \(I_{\rm sp}\) be the total information of the simultaneous receiver
and define the raw temporal information gain
\begin{equation}
g=\frac{I_{\rm t}}{I_{\rm sp}}.
\label{eq:information_gain_factor}
\end{equation}
The simultaneous benchmark is
\(D_{\rm sp}\simeq1/I_{\rm sp}\). A value \(g>1\) must correspond to a
real resource advantage, such as detector reuse, reduced optical
splitting, or longer integration per sequential setting; \(g\) is not an
intrinsic property of temporal readout.

Assuming the local measurement-noise error is unbiased and its leading
cross term with the mixed-time error is negligible, the fixed and ideal
adaptive temporal benchmarks are
\begin{subequations}
\label{eq:temporal_resource_mse}
\begin{align}
D_{\rm t,fixed}
&\simeq \frac{1}{gI_{\rm sp}}+\frac{3}{8}Q_\tau,\\
D_{\rm t,ad}
&\simeq \frac{1}{gI_{\rm sp}}
+\left(\frac{3}{8}-\frac{1}{\pi}\right)Q_\tau .
\end{align}
\end{subequations}
Defining \(\chi_Q=I_{\rm sp}Q_\tau\), temporal readout is favored when
\begin{subequations}
\label{eq:Q_boundaries}
\begin{align}
\chi_Q
&<\frac{8}{3}\left(1-\frac{1}{g}\right)
&&\text{(fixed)},\\
\chi_Q
&<\frac{1-1/g}{3/8-1/\pi}
&&\text{(adaptive)},
\end{align}
\end{subequations}
with \(g>1\).

For the increment-aware estimator, the relevant local uncertainty is
\(V_\tau=\Var(w_\tau\mid\mathcal H_0)\). In the symmetric benchmark,
each temporal quadrature has local information
\(J=gI_{\rm sp}/2\), giving
\[
D_{\rm t,IA}
=
\left[
J+\frac{J}{1+JV_\tau}
\right]^{-1}.
\]
With \(\chi_V=I_{\rm sp}V_\tau\), the condition
\(D_{\rm t,IA}<D_{\rm sp}\) becomes
\begin{equation}
\chi_V
<
\frac{4(g-1)}{g(2-g)},
\qquad
1<g<2.
\label{eq:ia_information_boundary}
\end{equation}
If \(V_\tau\) varies with predictor state, this is a local benchmark and
architecture-level performance requires averaging over those states.
For unequal quadrature information, the corresponding \(C\rightarrow S\)
criterion is
\begin{equation}
I_S^{(\rm t)}
+
\frac{I_C^{(\rm t)}}{1+I_C^{(\rm t)}V_\tau}
>
I_C^{(\rm sp)}+I_S^{(\rm sp)}.
\label{eq:general_crossover}
\end{equation}

\begin{figure*}[t]
\centering
\includegraphics[
    width=0.95\textwidth
]{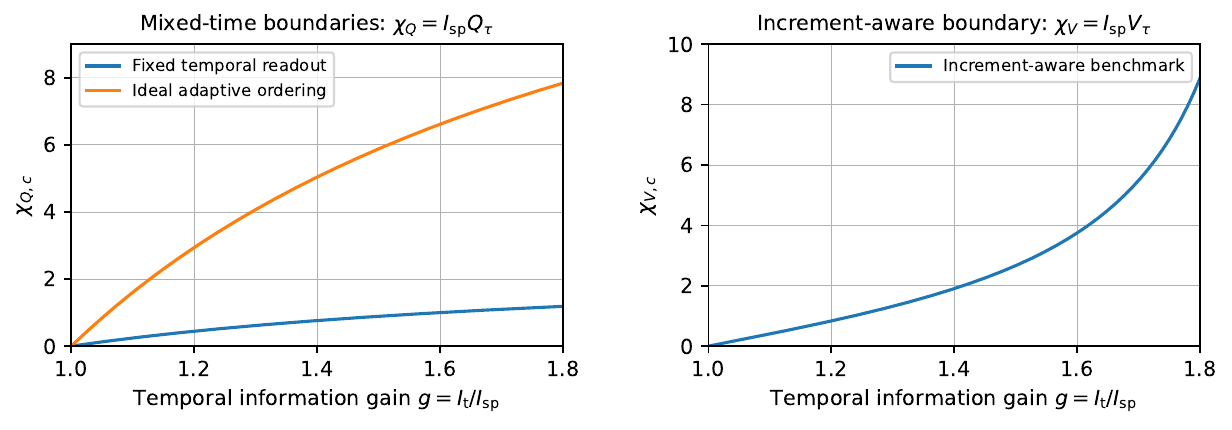}
\caption{
Dimensionless spatial--temporal architecture boundaries as a function
of the temporal information-gain factor
\(g=I_{\rm t}/I_{\rm sp}\).
(a) Fixed-order and ideal adaptive mixed-time boundaries, expressed in
terms of the total-increment parameter
\(\chi_Q=I_{\rm sp}Q_\tau\).
(b) Increment-aware boundary, expressed in terms of the residual
uncertainty parameter
\(\chi_V=I_{\rm sp}V_\tau\),
where \(V_\tau\) is the unresolved inter-measurement uncertainty after
the predictable phase evolution has been propagated.
Because \(Q_\tau\) and \(V_\tau\) represent different physical
quantities, the two panels should not be interpreted as sharing a
common vertical phase-evolution variable unless an explicit relation
between them is imposed.
}
\label{fig:architecture_boundaries}
\end{figure*}

The \(Q_\tau\)- and \(V_\tau\)-based boundaries should not be placed on a
single phase-evolution axis unless an explicit relation between
\(Q_\tau\) and \(V_\tau\) is supplied. The Fisher-information benchmark
and the algebra leading to the architecture boundaries are given in
Appendix~\ref{app:resource_derivation}. The present four-detector
experiment validates the mixed-time mechanism under a common photon
record; it does not demonstrate \(g>1\) or an experimental architecture
crossover.
\section{Discussion}
\label{sec:discussion}

The experiment validates the mixed-time mechanism at both the local and
finite-increment levels. The fitted first-order coefficients follow the
predicted \(-\sin^2\phi_0\) and \(-\cos^2\phi_0\) dependences with
correlations above \(0.998\), demonstrating that the temporal error is
strongly phase dependent rather than being described only by an averaged
drift penalty. The exact nonlinear model additionally reproduces the
pointwise same-record circular-error sequences with correlations near
\(0.992\), showing that the trajectory-level description remains accurate
beyond the perturbative regime.

An important feature of the formulation is that no zero-mean increment
assumption is required. The experimentally obtained
\(\widehat{Q}_{5\,\mathrm{ms}}=0.122401~\mathrm{rad}^2\) is the raw second
moment of the reconstructed phase increment rather than its variance.
Consequently, deterministic or systematic phase evolution contributes to
the mixed-time penalty through the squared increment in the same manner as
stochastic evolution. The exact error relations therefore remain
applicable without assigning the measured phase trajectory to a specific
Brownian, Ornstein--Uhlenbeck, or other stochastic model.

The complementary phase dependence of the two acquisition orders provides
the basis for phase-predicted ordering. In the small-increment,
phase-independent-\(Q_\tau\), uniform-phase limit, this strategy reduces
the leading mean-square coefficient from \(3/8\) to
\(3/8-1/\pi\), corresponding to an \(84.9\%\) reduction. This value is a
special analytical limit rather than a universal finite-increment
performance figure. Indeed, replacing uniform phase averaging by the
measured phase moments changes the leading-order prediction only slightly,
to \(84.73\%\), whereas evaluating the general exact nonlinear theory for
the measured phase trajectory gives a \(72.11\%\) reduction. The measured
same-record reduction is \(71.51\%\). Their agreement to within
\(0.60\) percentage points shows that the lower experimental reduction is
predominantly accounted for by finite-increment nonlinearities rather than
by nonuniform phase coverage alone.

The ordering rule also remains useful when the phase used to select the
acquisition order is imperfect. A one-cycle-stale phase predictor agrees
with the ideal ordering decision for \(82.80\%\) of the samples and yields
a \(47.17\%\) reduction in the same-record mean-square discrepancy.
Predictor error is most consequential near the switching boundaries,
where the preferred acquisition order changes. Away from these
boundaries, moderate phase-estimation error can leave the selected order
unchanged. A physical adaptive implementation therefore need not predict
the phase with uniform precision over the full phase range; the required
accuracy depends strongly on distance from the ordering boundaries.

The mixed-time and increment-aware descriptions address distinct aspects
of temporal acquisition. The quantity
\(Q_\tau=\E[\delta_\tau^2]\) measures the total mean-square phase evolution
occurring between the two quadrature measurements and governs the direct
mixed-time reconstruction penalty. By contrast,
\(V_\tau=\Var(w_\tau\mid\mathcal H_0)\) represents only the residual
uncertainty after predictable phase evolution has been propagated.
Accordingly, \(V_\tau\) determines the degradation of stale measurement
information through
\(J_{\rm eff}=J/(1+JV_\tau)\). The associated dimensionless parameters
\(\chi_Q=I_{\rm sp}Q_\tau\) and
\(\chi_V=I_{\rm sp}V_\tau\) therefore characterize different
spatial--temporal tradeoffs and should not be placed on a common
phase-evolution scale unless a dynamical model explicitly relates
\(Q_\tau\) and \(V_\tau\).

The same-record experimental design is both a strength and a limitation.
Because all four detector channels are recorded simultaneously, the
temporal estimates can be reconstructed from exactly the same underlying
photon record as the simultaneous reference. This removes switching
transients, configuration-dependent loss, and run-to-run phase
differences, allowing the loss of quadrature simultaneity to be isolated.
However, the simultaneous phase is itself a measured reference rather than
independently known ground truth, and
\(\widehat{Q}_\tau\) contains finite-count phase-reconstruction noise.
Furthermore, the experiment does not constitute an end-to-end
demonstration of a physically switched temporal receiver. Such a receiver
may introduce additional switching and settling errors, but it may also
provide a genuine hardware or information advantage \(g>1\). The resource
analysis should therefore be interpreted as an architecture-level design
benchmark rather than as an experimentally demonstrated crossover. These
considerations coexist with calibration and hardware-imperfection
constraints studied in programmable photonics
\cite{Bandyopadhyay2021,Xu2022,Litvin2025Calibration,Elmas2026Variance}.

Finally, the mixed-time mechanism is not specific to the present
programmable processor. Whenever a phase or related parameter is
reconstructed from complementary observables sampled at different times,
temporal evolution between those samples can produce the same class of
mixed-time error. The analysis is therefore also relevant to
phase-shifting and temporally interleaved interferometric receivers
\cite{Cheng2018,Roztocki2021}.

\section{Conclusion}
\label{sec:conclusion}

We have developed a general description of the error produced when
complementary optical quadratures are acquired at different times.
The exact circular end-time error depends only on the starting phase and
the realized inter-measurement phase increment and therefore does not
require a prescribed stochastic phase-noise model. In the
small-increment regime, the two acquisition orders obey
\(e_{C\rightarrow S}\simeq-\delta_\tau\sin^2\phi_0\) and
\(e_{S\rightarrow C}\simeq-\delta_\tau\cos^2\phi_0\), revealing
complementary phase sensitivities and motivating phase-predicted
quadrature ordering.

A continuous 30-min four-channel photon-count experiment provides a
controlled same-record validation of this mixed-time mechanism. At
\(\tau=\SI{5}{\milli\second}\), the first-order phase dependence agrees
with theory at correlations above \(0.998\), while the exact nonlinear
prediction reproduces the pointwise circular errors with correlations
near \(0.992\). Applying the phase-predicted ordering rule to the measured
record reduces the fixed-order mean-square discrepancy by \(71.51\%\);
the exact theory evaluated for the same measured phase trajectory predicts
\(72.11\%\). The commonly quoted \(84.9\%\) reduction is recovered only
as the additional uniform-phase, phase-independent-\(Q_\tau\),
small-increment limit.

Finally, the total increment statistic \(Q_\tau\) and the residual
prediction uncertainty \(V_\tau\) play fundamentally different roles.
The former governs direct mixed-time reconstruction error, whereas the
latter determines the information retained by a propagated stale
measurement through
\(J_{\rm eff}=J/(1+JV_\tau)\). This distinction provides a basis for
comparing simultaneous and resource-reduced temporal receivers while
separating unavoidable mixed-time error from phase evolution that can, in
principle, be predicted and compensated.
\section*{AUTHOR DECLARATIONS}

\subsection*{Conflict of Interest}

The authors have no conflicts to disclose.

% Complete after confirming the contribution roles of all authors.
\section*{DATA AVAILABILITY}

The data and analysis code that support the findings of this study are
available from the corresponding author upon reasonable request.
\appendices

\section{Exact and Perturbative Mixed-Time Error}
\label{app:mixed_time_derivation}

This appendix gives the intermediate steps behind the mixed-time error
laws used in the main text. Let
\begin{equation}
\phi_1=\phi_0+\delta_\tau,
\end{equation}
and define the estimation target as \(\phi_1\).

\subsection{Cosine-first ordering}

For \(C\rightarrow S\),
\begin{equation}
\widehat{\phi}_{C\rightarrow S}
=
\atanTwo\!\left(\sin\phi_1,\cos\phi_0\right).
\end{equation}
Writing
\begin{equation}
z_{C\rightarrow S}
=
\cos\phi_0+\ii\sin\phi_1,
\end{equation}
the circular end-time error can be written without subtracting two
branch-dependent angles:
\begin{equation}
e_{C\rightarrow S}
=
\operatorname{Arg}
\!\left(
z_{C\rightarrow S}e^{-\ii\phi_1}
\right).
\end{equation}
Hence,
\begin{equation}
e_{C\rightarrow S}
=
\atanTwo
\!\left(
Y_{C\rightarrow S},
X_{C\rightarrow S}
\right),
\label{eq:app_cs_exact}
\end{equation}
with
\begin{subequations}
\begin{align}
Y_{C\rightarrow S}
&=
\sin\phi_1
\left(
\cos\phi_1-\cos\phi_0
\right),\\
X_{C\rightarrow S}
&=
\cos\phi_0\cos\phi_1+\sin^2\phi_1.
\end{align}
\end{subequations}

For the first-order expansion, use
\begin{equation}
d\,\atanTwo(y,x)
=
\frac{x\,dy-y\,dx}{x^2+y^2}.
\label{eq:app_atan2_diff}
\end{equation}
Here,
\begin{equation}
x=\cos\phi_0,
\qquad
y=\sin(\phi_0+\delta_\tau).
\end{equation}
At \(\delta_\tau=0\),
\begin{equation}
x^2+y^2=1,
\qquad
dx=0,
\qquad
dy=\cos\phi_0\,d\delta_\tau,
\end{equation}
so that
\begin{equation}
\left.
\frac{\partial\widehat{\phi}_{C\rightarrow S}}
{\partial\delta_\tau}
\right|_{\delta_\tau=0}
=
\cos^2\phi_0.
\end{equation}
Therefore,
\begin{equation}
\widehat{\phi}_{C\rightarrow S}
=
\phi_0+\delta_\tau\cos^2\phi_0
+\mathcal{O}(\delta_\tau^2),
\end{equation}
and subtraction of the end-time target
\(\phi_1=\phi_0+\delta_\tau\) gives
\begin{equation}
e_{C\rightarrow S}
=
-\delta_\tau\sin^2\phi_0
+\mathcal{O}(\delta_\tau^2).
\label{eq:app_cs_first}
\end{equation}

Expanding Eq.~\eqref{eq:app_cs_exact} one order further gives
\begin{equation}
\begin{split}
e_{C\rightarrow S}
={}&
-\delta_\tau\sin^2\phi_0\\
&-
\delta_\tau^2
\left[
\frac{1}{2}\sin(2\phi_0)
+
\frac{1}{8}\sin(4\phi_0)
\right]
+\mathcal{O}(\delta_\tau^3).
\end{split}
\label{eq:app_cs_second}
\end{equation}

\subsection{Sine-first ordering}

For \(S\rightarrow C\),
\begin{equation}
\widehat{\phi}_{S\rightarrow C}
=
\atanTwo\!\left(\sin\phi_0,\cos\phi_1\right).
\end{equation}
Writing
\begin{equation}
z_{S\rightarrow C}
=
\cos\phi_1+\ii\sin\phi_0,
\end{equation}
the exact circular error is
\begin{equation}
e_{S\rightarrow C}
=
\atanTwo
\!\left(
Y_{S\rightarrow C},
X_{S\rightarrow C}
\right),
\label{eq:app_sc_exact}
\end{equation}
where
\begin{subequations}
\begin{align}
Y_{S\rightarrow C}
&=
\cos\phi_1
\left(
\sin\phi_0-\sin\phi_1
\right),\\
X_{S\rightarrow C}
&=
\cos^2\phi_1+\sin\phi_0\sin\phi_1.
\end{align}
\end{subequations}

Using Eq.~\eqref{eq:app_atan2_diff} with
\begin{equation}
x=\cos(\phi_0+\delta_\tau),
\qquad
y=\sin\phi_0,
\end{equation}
gives at \(\delta_\tau=0\)
\begin{equation}
dx=-\sin\phi_0\,d\delta_\tau,
\qquad
dy=0,
\end{equation}
and hence
\begin{equation}
\left.
\frac{\partial\widehat{\phi}_{S\rightarrow C}}
{\partial\delta_\tau}
\right|_{\delta_\tau=0}
=
\sin^2\phi_0.
\end{equation}
Thus,
\begin{equation}
e_{S\rightarrow C}
=
-\delta_\tau\cos^2\phi_0
+\mathcal{O}(\delta_\tau^2).
\label{eq:app_sc_first}
\end{equation}
The second-order expansion is
\begin{equation}
\begin{split}
e_{S\rightarrow C}
={}&
-\delta_\tau\cos^2\phi_0\\
&+
\delta_\tau^2
\left[
\frac{1}{2}\sin(2\phi_0)
-
\frac{1}{8}\sin(4\phi_0)
\right]
+\mathcal{O}(\delta_\tau^3).
\end{split}
\label{eq:app_sc_second}
\end{equation}

\section{Mean-Square and Adaptive Coefficients}
\label{app:mse_adaptive_derivation}

Squaring Eqs.~\eqref{eq:app_cs_first} and
\eqref{eq:app_sc_first} gives, to leading order,
\begin{subequations}
\begin{align}
D_{C\rightarrow S}
&=
\E\!\left[
\delta_\tau^2\sin^4\phi_0
\right]
+
\mathcal{O}\!\left(\E[|\delta_\tau|^3]\right),\\
D_{S\rightarrow C}
&=
\E\!\left[
\delta_\tau^2\cos^4\phi_0
\right]
+
\mathcal{O}\!\left(\E[|\delta_\tau|^3]\right).
\end{align}
\end{subequations}

Defining the conditional increment second moment
\begin{equation}
q_\tau(\phi)
=
\E\!\left[
\delta_\tau^2
\mid
\phi_0=\phi
\right],
\end{equation}
the same result may be written as
\begin{subequations}
\begin{align}
D_{C\rightarrow S}
&=
\int
q_\tau(\phi)\sin^4\phi\,
p_\phi(\phi)\,d\phi+\cdots,\\
D_{S\rightarrow C}
&=
\int
q_\tau(\phi)\cos^4\phi\,
p_\phi(\phi)\,d\phi+\cdots.
\end{align}
\end{subequations}
These are the general first-order mean-square laws and do not require
independence between \(\phi_0\) and \(\delta_\tau\).

If
\begin{equation}
q_\tau(\phi)=Q_\tau
\end{equation}
and \(\phi_0\) is uniform on \([0,2\pi)\), then
\begin{equation}
\E[\sin^4\phi]
=
\E[\cos^4\phi]
=
\frac{3}{8},
\end{equation}
which gives
\begin{equation}
D_{\rm fixed}
\simeq
\frac{3}{8}Q_\tau.
\end{equation}

For ideal adaptive ordering, the smaller of the two leading squared
errors is selected:
\begin{equation}
D_{\rm ad}
\simeq
\E\!\left[
\delta_\tau^2
\min(\sin^4\phi_0,\cos^4\phi_0)
\right].
\end{equation}
Under the same phase-independent-\(Q_\tau\) and uniform-phase
conditions,
\begin{align}
\E\!\left[
\min(\sin^4\phi,\cos^4\phi)
\right]
&=
\frac{8}{2\pi}
\int_0^{\pi/4}
\sin^4\phi\,d\phi
\nonumber\\
&=
\frac{4}{\pi}
\left[
\frac{3\phi}{8}
-\frac{\sin(2\phi)}{4}
+\frac{\sin(4\phi)}{32}
\right]_{0}^{\pi/4}
\nonumber\\
&=
\frac{3}{8}-\frac{1}{\pi}.
\end{align}
Therefore,
\begin{equation}
D_{\rm ad}
\simeq
\left(
\frac{3}{8}-\frac{1}{\pi}
\right)
Q_\tau.
\end{equation}
The asymptotic adaptive-to-fixed ratio is
\begin{equation}
\frac{D_{\rm ad}}{D_{\rm fixed}}
=
\frac{3/8-1/\pi}{3/8}
=
1-\frac{8}{3\pi}
\simeq
0.1512,
\end{equation}
so the corresponding reduction in leading MSE is
\begin{equation}
\frac{8}{3\pi}
\simeq
84.9\%.
\end{equation}

This result is the uniform-phase, small-increment limit of the general
exact performance measures in
Eqs.~\eqref{eq:general_exact_fixed_mse}--\eqref{eq:exact_adaptive_reduction_general}.
For finite increments or nonuniform joint phase statistics, those exact
expressions should be evaluated directly rather than replacing them by
the \(84.9\%\) special-case value. Section~\ref{sec:experimental_validation}
performs this exact evaluation for the experimental phase evolution and
compares it with the measured adaptive reduction.

\section{Increment-Aware Estimator Derivation}
\label{app:increment_aware_derivation}

Consider first a noisy cosine observation
\begin{equation}
\widetilde{C}
=
\cos\phi+n_C,
\qquad
\Var(n_C)=v_C.
\end{equation}
Linearizing around a predictor \(\bar{\phi}\),
\begin{equation}
\widetilde{C}
\simeq
\cos\bar{\phi}
-
\sin\bar{\phi}(\phi-\bar{\phi})
+n_C.
\end{equation}
The corresponding local phase observation is
\begin{equation}
z_C
=
\bar{\phi}
-
\frac{
\widetilde{C}-\cos\bar{\phi}
}{
\sin\bar{\phi}
}
\simeq
\phi+\epsilon_C,
\end{equation}
with
\begin{equation}
\Var(\epsilon_C)
=
\frac{v_C}{\sin^2\bar{\phi}}.
\end{equation}
Thus,
\begin{equation}
J_C
=
\frac{\sin^2\bar{\phi}}{v_C}.
\end{equation}
The analogous sine observation gives
\begin{equation}
J_S
=
\frac{\cos^2\bar{\phi}}{v_S}.
\end{equation}

Let
\begin{equation}
m_\tau
=
\E[\delta_\tau\mid\mathcal{H}_0],
\qquad
w_\tau
=
\delta_\tau-m_\tau,
\end{equation}
with
\begin{equation}
\E[w_\tau\mid\mathcal{H}_0]=0,
\qquad
\Var(w_\tau\mid\mathcal{H}_0)=V_\tau.
\end{equation}
For \(C\rightarrow S\), write
\begin{equation}
z_{C,0}
=
\phi_0+\epsilon_C,
\qquad
z_{S,1}
=
\phi_1+\epsilon_S.
\end{equation}
Propagating the stale cosine observation by the predicted increment,
\begin{align}
z_{C,0}^{\rightarrow1}
&=
z_{C,0}+m_\tau
\nonumber\\
&=
\phi_1+\epsilon_C-w_\tau.
\end{align}
Assuming the local measurement error and the residual phase uncertainty
are independent within the estimator model,
\begin{equation}
\Var
\!\left(
z_{C,0}^{\rightarrow1}-\phi_1
\mid\mathcal{H}_0
\right)
=
J_C^{-1}+V_\tau.
\end{equation}
The propagated stale information is therefore
\begin{equation}
J_{C,\rm eff}
=
\frac{1}{J_C^{-1}+V_\tau}
=
\frac{J_C}{1+J_CV_\tau}.
\end{equation}
More generally,
\begin{equation}
J_{\rm eff}
=
\frac{J}{1+JV_\tau}.
\end{equation}

The minimum-variance local combination of the fresh and propagated stale
observations is the inverse-variance weighted estimate
\begin{equation}
\widehat{\phi}_{1,C\rightarrow S}^{\rm IA}
=
\frac{
J_S z_{S,1}
+
J_{C,\rm eff}z_{C,0}^{\rightarrow1}
}{
J_S+J_{C,\rm eff}
},
\end{equation}
whose local MSE is
\begin{equation}
D_{C\rightarrow S}^{\rm IA}
=
\left[
J_S
+
\frac{J_C}{1+J_CV_\tau}
\right]^{-1}.
\end{equation}
The \(S\rightarrow C\) expression follows by exchanging \(C\) and \(S\).

The loss of information caused by staleness is
\begin{equation}
\Delta J
=
J-J_{\rm eff}
=
\frac{J^2V_\tau}{1+JV_\tau},
\end{equation}
and
\begin{equation}
\frac{\partial\Delta J}{\partial J}
=
\frac{
JV_\tau(2+JV_\tau)
}{
(1+JV_\tau)^2
}
>0.
\end{equation}
Thus, delaying a more informative quadrature produces the larger
information loss, reproducing the adaptive ordering rule.

Finally,
\begin{equation}
\delta_\tau=m_\tau+w_\tau
\end{equation}
implies
\begin{equation}
\E[\delta_\tau^2\mid\mathcal{H}_0]
=
m_\tau^2+V_\tau,
\end{equation}
and averaging over predictor states gives
\begin{equation}
Q_\tau
=
\E[\delta_\tau^2]
=
\E[m_\tau^2+V_\tau].
\end{equation}
This explicitly separates total mean-square phase evolution from
unresolved prediction uncertainty.

\section{Resource-Boundary Derivations}
\label{app:resource_derivation}

For an ideal complementary cosine pair, let
\begin{subequations}
\begin{align}
\lambda_{+}(\phi)
&=
\frac{\eta\Lambda}{2}(1+\cos\phi),\\
\lambda_{-}(\phi)
&=
\frac{\eta\Lambda}{2}(1-\cos\phi).
\end{align}
\end{subequations}
For independent Poisson counts, the Fisher information is
\begin{align}
I(\phi)
&=
\sum_{k\in\{+,-\}}
\frac{1}{\lambda_k(\phi)}
\left(
\frac{\partial\lambda_k}{\partial\phi}
\right)^2
\nonumber\\
&=
\frac{\eta\Lambda}{2}\sin^2\phi
\left[
\frac{1}{1+\cos\phi}
+
\frac{1}{1-\cos\phi}
\right]
\nonumber\\
&=
\eta\Lambda.
\end{align}
The sine pair gives the same result.

Let \(I_{\rm sp}\) denote the total simultaneous-receiver information and
\(I_{\rm t}=gI_{\rm sp}\) the raw temporal information. The simultaneous
benchmark is
\begin{equation}
D_{\rm sp}
\simeq
\frac{1}{I_{\rm sp}}.
\end{equation}
Assuming that the local measurement-noise error is unbiased and its
leading cross term with the mixed-time error is negligible, write
\begin{equation}
D_{\rm t}
\simeq
\frac{1}{gI_{\rm sp}}+aQ_\tau,
\end{equation}
where
\begin{equation}
a_{\rm fixed}=\frac{3}{8},
\qquad
a_{\rm ad}=\frac{3}{8}-\frac{1}{\pi}.
\end{equation}
The temporal architecture is favored when
\begin{equation}
\frac{1}{gI_{\rm sp}}+aQ_\tau
<
\frac{1}{I_{\rm sp}}.
\end{equation}
Multiplication by \(I_{\rm sp}\) gives
\begin{equation}
\frac{1}{g}+a\chi_Q<1,
\qquad
\chi_Q=I_{\rm sp}Q_\tau,
\end{equation}
or
\begin{equation}
\chi_Q
<
\frac{1-1/g}{a}.
\end{equation}
Hence,
\begin{subequations}
\begin{align}
\chi_Q
&<
\frac{8}{3}
\left(
1-\frac{1}{g}
\right)
&&\text{(fixed)},\\
\chi_Q
&<
\frac{1-1/g}{3/8-1/\pi}
&&\text{(adaptive)}.
\end{align}
\end{subequations}

For the symmetric increment-aware benchmark,
\begin{equation}
J
=
\frac{gI_{\rm sp}}{2}.
\end{equation}
The total end-time information is
\begin{equation}
J_{\rm tot}
=
J+\frac{J}{1+JV_\tau},
\end{equation}
so that
\begin{equation}
D_{\rm t,IA}
=
\frac{1}{J_{\rm tot}}.
\end{equation}
The requirement
\begin{equation}
D_{\rm t,IA}
<
\frac{1}{I_{\rm sp}}
\end{equation}
is equivalent to
\begin{equation}
\frac{g}{2}
\left[
1+
\frac{1}{
1+\dfrac{g}{2}\chi_V
}
\right]
>1,
\qquad
\chi_V=I_{\rm sp}V_\tau.
\end{equation}
For \(1<g<2\), solving for \(\chi_V\) gives
\begin{equation}
\chi_V
<
\frac{4(g-1)}{g(2-g)}.
\end{equation}
At \(g=2\), the fresh quadrature alone carries the full spatial
information benchmark; for \(g>2\), it exceeds that benchmark under the
assumed resource convention.

For unequal temporal quadrature informations, the corresponding
\(C\rightarrow S\) condition is simply
\begin{equation}
I_S^{(\rm t)}
+
\frac{
I_C^{(\rm t)}
}{
1+I_C^{(\rm t)}V_\tau
}
>
I_C^{(\rm sp)}
+
I_S^{(\rm sp)},
\end{equation}
with \(C\) and \(S\) exchanged for the reverse order.

\bibliographystyle{IEEEtran}
\bibliography{references}

@article{Clements2016,
  author  = {Clements, William R. and Humphreys, Peter C. and
             Metcalf, Benjamin J. and Kolthammer, W. Steven and
             Walmsley, Ian A.},
  title   = {Optimal Design for Universal Multiport Interferometers},
  journal = {Optica},
  volume  = {3},
  number  = {12},
  pages   = {1460--1465},
  year    = {2016},
  doi     = {10.1364/OPTICA.3.001460}
}

@article{Harris2018,
  author  = {Harris, Nicholas C. and Carolan, Jacques and Bunandar, Darius and
             Prabhu, Mihika and Hochberg, Michael and Baehr-Jones, Tom and
             Fanto, Michael L. and Smith, A. Matthew and Tison, Christopher C.
             and Alsing, Paul M. and Englund, Dirk},
  title   = {Linear Programmable Nanophotonic Processors},
  journal = {Optica},
  volume  = {5},
  number  = {12},
  pages   = {1623--1631},
  year    = {2018},
  doi     = {10.1364/OPTICA.5.001623}
}

@article{Bogaerts2020,
  author  = {Bogaerts, Wim and P{\'e}rez, Daniel and Capmany, Jos{\'e} and
             Miller, David A. B. and Poon, Joyce and Englund, Dirk and
             Morichetti, Francesco and Melloni, Andrea},
  title   = {Programmable Photonic Circuits},
  journal = {Nature},
  volume  = {586},
  pages   = {207--216},
  year    = {2020},
  doi     = {10.1038/s41586-020-2764-0}
}

@article{Elmas2025Phase,
  author  = {Elmas, G{\"o}khan and Litvin, Igor A. and Kohl, Paul and
             N{\"o}tzel, Janis},
  title   = {Modeling and Analysis of Phase Instability in a Photonic Processor},
  journal = {Applied Optics},
  volume  = {64},
  number  = {14},
  pages   = {3995--4003},
  year    = {2025},
  doi     = {10.1364/AO.560370}
}

@article{Kinnstaetter1988,
  author  = {Kinnstaetter, K. and Lohmann, Adolf W. and
             Schwider, Johannes and Streibl, Norbert},
  title   = {Accuracy of Phase Shifting Interferometry},
  journal = {Applied Optics},
  volume  = {27},
  number  = {24},
  pages   = {5082--5089},
  year    = {1988},
  doi     = {10.1364/AO.27.005082}
}

@article{Deck1998,
  author  = {Deck, Leslie L. and de Groot, Peter J.},
  title   = {Punctuated Quadrature Phase-Shifting Interferometry},
  journal = {Optics Letters},
  volume  = {23},
  number  = {1},
  pages   = {19--21},
  year    = {1998},
  doi     = {10.1364/OL.23.000019}
}

@article{Roztocki2021,
  author  = {Roztocki, Piotr and MacLellan, Benjamin and Islam, Mehedi and
             Reimer, Christian and Fischer, Bennet and Sciara, Stefania and
             Helsten, Robin and Jestin, Yoann and Cino, Alfonso and
             Chu, Sai T. and Little, Brent and Moss, David J. and
             Kues, Michael and Morandotti, Roberto},
  title   = {Arbitrary Phase Access for Stable Fiber Interferometers},
  journal = {Laser \& Photonics Reviews},
  volume  = {15},
  number  = {7},
  pages   = {2000524},
  year    = {2021},
  doi     = {10.1002/lpor.202000524}
}

@article{Wiseman1995,
  author  = {Wiseman, H. M.},
  title   = {Adaptive Phase Measurements of Optical Modes:
             Going Beyond the Marginal {Q} Distribution},
  journal = {Physical Review Letters},
  volume  = {75},
  number  = {25},
  pages   = {4587--4590},
  year    = {1995},
  doi     = {10.1103/PhysRevLett.75.4587}
}

@article{BerryWiseman2000,
  author  = {Berry, D. W. and Wiseman, H. M.},
  title   = {Optimal States and Almost Optimal Adaptive Measurements
             for Quantum Interferometry},
  journal = {Physical Review Letters},
  volume  = {85},
  number  = {24},
  pages   = {5098--5101},
  year    = {2000},
  doi     = {10.1103/PhysRevLett.85.5098}
}

@book{Kay1993,
  author    = {Kay, Steven M.},
  title     = {Fundamentals of Statistical Signal Processing:
               Estimation Theory},
  volume    = {1},
  publisher = {Prentice Hall},
  year      = {1993},
  isbn      = {9780133457117}
}

@book{Sarkka2013,
  author    = {S{\"a}rkk{\"a}, Simo},
  title     = {Bayesian Filtering and Smoothing},
  publisher = {Cambridge University Press},
  year      = {2013},
  doi       = {10.1017/CBO9781139344203}
}

@article{Miller2013,
  author  = {Miller, David A. B.},
  title   = {Self-Configuring Universal Linear Optical Component},
  journal = {Photonics Research},
  volume  = {1},
  number  = {1},
  pages   = {1--15},
  year    = {2013},
  doi     = {10.1364/PRJ.1.000001}
}

@article{Perez2017,
  author  = {P{\'e}rez, Daniel and Gasulla, Ivana and Crudgington, Lee and
             Thomson, David J. and Khokhar, Ali Z. and Li, Ke and Cao, Wei and
             Mashanovich, Goran Z. and Capmany, Jos{\'e}},
  title   = {Multipurpose Silicon Photonics Signal Processor Core},
  journal = {Nature Communications},
  volume  = {8},
  pages   = {636},
  year    = {2017},
  doi     = {10.1038/s41467-017-00714-1}
}

@article{PerezLopez2020,
  author  = {P{\'e}rez-L{\'o}pez, Daniel and L{\'o}pez, Aitor and
             DasMahapatra, Prometheus and Capmany, Jos{\'e}},
  title   = {Multipurpose Self-Configuration of Programmable Photonic Circuits},
  journal = {Nature Communications},
  volume  = {11},
  pages   = {6359},
  year    = {2020},
  doi     = {10.1038/s41467-020-19608-w}
}

@article{Annoni2017,
  author  = {Annoni, Andrea and Guglielmi, Emanuele and Carminati, Marco and
             Ferrari, Giorgio and Sampietro, Marco and Miller, David A. B. and
             Melloni, Andrea and Morichetti, Francesco},
  title   = {Unscrambling Light---Automatically Undoing Strong Mixing Between Modes},
  journal = {Light: Science \& Applications},
  volume  = {6},
  pages   = {e17110},
  year    = {2017},
  doi     = {10.1038/lsa.2017.110}
}

@article{Carolan2015,
  author  = {Carolan, Jacques and Harrold, Christopher and Sparrow, Chris and
             Mart{\'i}n-L{\'o}pez, Enrique and Russell, Nicholas J. and
             Silverstone, Joshua W. and Shadbolt, Peter J. and Matsuda, Nobuyuki and
             Oguma, Manabu and Itoh, Mikitaka and Marshall, Graham D. and
             Thompson, Mark G. and Matthews, Jonathan C. F. and Hashimoto, Toshikazu and
             O'Brien, Jeremy L. and Laing, Anthony},
  title   = {Universal Linear Optics},
  journal = {Science},
  volume  = {349},
  number  = {6249},
  pages   = {711--716},
  year    = {2015},
  doi     = {10.1126/science.aab3642}
}

@article{Taballione2019,
  author  = {Taballione, Caterina and Wolterink, Tom A. W. and Lugani, Jasleen and
             Eckstein, Andreas and Bell, Bryn A. and Grootjans, Robert and
             Visscher, Ilka and Geskus, Dimitri and Roeloffzen, Chris G. H. and
             Renema, Jelmer J. and Walmsley, Ian A. and Pinkse, Pepijn W. H. and
             Boller, Klaus-J.},
  title   = {8x8 Reconfigurable Quantum Photonic Processor Based on Silicon Nitride Waveguides},
  journal = {Optics Express},
  volume  = {27},
  number  = {19},
  pages   = {26842--26857},
  year    = {2019},
  doi     = {10.1364/OE.27.026842}
}

@article{Taballione2021,
  author  = {Taballione, Caterina and van der Meer, Reinier and Snijders, Henk J. and
             Hooijschuur, Peter and Epping, J{\"o}rn P. and de Goede, Michiel and
             Kassenberg, Ben and Venderbosch, Pim and Toebes, Chris and
             van den Vlekkert, Hans and Pinkse, Pepijn W. H. and Renema, Jelmer J.},
  title   = {A Universal Fully Reconfigurable 12-Mode Quantum Photonic Processor},
  journal = {Materials for Quantum Technology},
  volume  = {1},
  number  = {3},
  pages   = {035002},
  year    = {2021},
  doi     = {10.1088/2633-4356/ac168c}
}

@article{Bandyopadhyay2021,
  author  = {Bandyopadhyay, Saumil and Hamerly, Ryan and Englund, Dirk},
  title   = {Hardware Error Correction for Programmable Photonics},
  journal = {Optica},
  volume  = {8},
  number  = {10},
  pages   = {1247--1255},
  year    = {2021},
  doi     = {10.1364/OPTICA.424052}
}

@article{Xu2022,
  author  = {Xu, Xingyuan and Ren, Guanghui and Feleppa, Tim and Liu, Xumeng and
             Boes, Andreas and Mitchell, Arnan and Lowery, Arthur J.},
  title   = {Self-Calibrating Programmable Photonic Integrated Circuits},
  journal = {Nature Photonics},
  volume  = {16},
  pages   = {595--602},
  year    = {2022},
  doi     = {10.1038/s41566-022-01020-z}
}

@article{Markowitz2023,
  author  = {Markowitz, Matthew and Zelaya, Kevin and Miri, Mohammad-Ali},
  title   = {Auto-Calibrating Universal Programmable Photonic Circuits:
             Hardware Error-Correction and Defect Resilience},
  journal = {Optics Express},
  volume  = {31},
  number  = {23},
  pages   = {37673--37682},
  year    = {2023},
  doi     = {10.1364/OE.502226}
}

@article{Zheng2024,
  author  = {Zheng, Jin-Hao and Wang, Qin-Qin and Feng, Lan-Tian and Ding, Yu-Yang and
             Xu, Xiao-Ye and Ren, Xi-Feng and Li, Chuan-Feng and Guo, Guang-Can},
  title   = {Global Calibration of Large-Scale Photonic Integrated Circuits},
  journal = {Physical Review Applied},
  volume  = {22},
  number  = {5},
  pages   = {054011},
  year    = {2024},
  doi     = {10.1103/PhysRevApplied.22.054011}
}

@article{deGroot1995,
  author  = {de Groot, Peter J.},
  title   = {Vibration in Phase-Shifting Interferometry},
  journal = {Journal of the Optical Society of America A},
  volume  = {12},
  number  = {2},
  pages   = {354--365},
  year    = {1995},
  doi     = {10.1364/JOSAA.12.000354}
}

@article{Deck2009,
  author  = {Deck, Leslie L.},
  title   = {Suppressing Phase Errors from Vibration in Phase-Shifting Interferometry},
  journal = {Applied Optics},
  volume  = {48},
  number  = {20},
  pages   = {3948--3960},
  year    = {2009},
  doi     = {10.1364/AO.48.003948}
}

@article{Cheng2018,
  author  = {Cheng, Zhongtao and Liu, Dong},
  title   = {Fast and Accurate Wavefront Reconstruction in Two-Frame Phase-Shifting
             Interferometry with Unknown Phase Step},
  journal = {Optics Letters},
  volume  = {43},
  number  = {13},
  pages   = {3033--3036},
  year    = {2018},
  doi     = {10.1364/OL.43.003033}
}

@article{Armen2002,
  author  = {Armen, Michael A. and Au, John K. and Stockton, John K. and
             Doherty, Andrew C. and Mabuchi, Hideo},
  title   = {Adaptive Homodyne Measurement of Optical Phase},
  journal = {Physical Review Letters},
  volume  = {89},
  number  = {13},
  pages   = {133602},
  year    = {2002},
  doi     = {10.1103/PhysRevLett.89.133602}
}

@article{Yonezawa2012,
  author  = {Yonezawa, Hidehiro and Nakane, Daisuke and Wheatley, Trevor A. and
             Iwasawa, Kohjiro and Takeda, Shuntaro and Arao, Hajime and
             Ohki, Kentaro and Tsumura, Koji and Berry, Dominic W. and
             Ralph, Timothy C. and Wiseman, Howard M. and Huntington, Elanor H. and
             Furusawa, Akira},
  title   = {Quantum-Enhanced Optical-Phase Tracking},
  journal = {Science},
  volume  = {337},
  number  = {6101},
  pages   = {1514--1517},
  year    = {2012},
  doi     = {10.1126/science.1225258}
}

@article{Berni2015,
  author  = {Berni, Adriano A. and Gehring, Tobias and Nielsen, Bo M. and
             H{\"a}ndchen, Vitus and Paris, Matteo G. A. and Andersen, Ulrik L.},
  title   = {Ab Initio Quantum-Enhanced Optical Phase Estimation Using Real-Time Feedback Control},
  journal = {Nature Photonics},
  volume  = {9},
  pages   = {577--581},
  year    = {2015},
  doi     = {10.1038/nphoton.2015.139}
}

@article{Litvin2025Stable,
  author  = {Litvin, Igor A. and Elmas, G{\"o}khan and Kohl, Paul and N{\"o}tzel, Janis},
  title   = {Stable Signal Processing With a Photonic Processor},
  journal = {Journal of Lightwave Technology},
  volume  = {43},
  number  = {22},
  pages   = {9981--9990},
  year    = {2025},
  doi     = {10.1109/JLT.2025.3606027}
}

@article{Litvin2025Calibration,
  author  = {Litvin, Igor A. and Elmas, G{\"o}khan and El-Safty, Kareem and
             Chaudhary, Sumit and N{\"o}tzel, Janis},
  title   = {Robust Calibration and Energy Optimization in Reconfigurable Photonic Processors},
  journal = {Optics Express},
  volume  = {33},
  number  = {16},
  pages   = {35011--35027},
  year    = {2025},
  doi     = {10.1364/OE.566817}
}

@article{Litvin2026Stabilization,
  author  = {Litvin, Igor A. and Elmas, G{\"o}khan and Kohl, Paul and N{\"o}tzel, Janis},
  title   = {Multi-Input Signal Phase Stabilization in Photonic Processors with On-Chip Feedback Control},
  journal = {Optics Express},
  volume  = {34},
  number  = {6},
  pages   = {11244--11258},
  year    = {2026},
  doi     = {10.1364/OE.579969}
}

@article{Litvin2026JointDetection,
  author  = {Litvin, Igor A. and Elmas, G{\"o}khan and N{\"o}tzel, Janis},
  title   = {Joint Detection on Reconfigurable Photonic Processors With Multi-Input Phase Stabilization},
  journal = {IEEE Photonics Journal},
  volume  = {18},
  number  = {3},
  pages   = {6600909:1--9},
  year    = {2026},
  doi     = {10.1109/JPHOT.2026.3683204}
}

@article{Elmas2026Variance,
  author  = {Gökhan Elmas and Igor A. Litvin and Janis Nötzel},
  title   = {Local Variance-Based Calibration of Programmable Photonic Interferometer Meshes},
  journal = {arXiv preprint arXiv:2607.14759},
  year    = {2026}
}

\end{document}